\documentclass{PoS}

\usepackage{amssymb}
\usepackage{hepunits}
\usepackage{slashed}
\usepackage{wasysym}
\usepackage{graphicx}
\usepackage{subfig}

\newcommand{\beq}{\begin{equation}}
\newcommand{\eeq}{\end{equation}}
\newcommand{\bfig}{\begin{figure}[ht]\begin{center}}
\newcommand{\efig}{\end{center}\end{figure}}

\newcommand{\rismom}{{\rm RI/SMOM}}

\newcommand{\rismomqslash}{{\rismom_{\slashed{q}}}}
\newcommand{\rismomgamma}{{\rismom_{\gamma_{\mu}}}}

\newcommand{\Nf}{N_{\mathrm f}}

\title{$\Nf=2+1+1$ renormalisation of four-quark operators}

\ShortTitle{Nf=2+1+1 renormalisation of four-quark operators}

\author{\speaker{Julien Frison}\\
        School of Physics and Astronomy, The University of Edinburgh, Edinburgh EH9 3FD, UK\\
        E-mail: \email{jfrison@ph.ed.ac.uk}}

\author{Peter Boyle\\
        School of Physics and Astronomy, The University of Edinburgh, Edinburgh EH9 3FD, UK\\
        E-mail: \email{pab@ph.ed.ac.uk}}

\author{Nicolas Garron\\
        School of Computing and Mathematics, Plymouth University, Plymouth PL4 8AA, UK\\
        E-mail: \email{nicolas.garron@plymouth.ac.uk}}

\author{RBC-UKQCD collaboration}

\abstract{When several four-quark operators are allowed to mix through renormalisation, this can considerably amplify the problems coming from perturbative truncation and
discretisation effects. In this work we investigate whether our previous $B_K$ strategy can conveniently be generalised to a wider set of operators, corresponding to the so-called
``SUSY $B_K$'' basis of four-quark operators. We show that the discretisation effects, when plotted as a function of $ap$, are surprisingly independent of the lattice spacing. They appear reasonably under control up to very large energy scales. This allows us to discuss the effect of varying the intermediate scale on which the perturbative matching is done, and therefore the prospects of future high-precision studies with a Rome-Southampton renormalisation. }

\FullConference{The 33rd International Symposium on Lattice Field Theory\\
                 14-18 July, 2015\\
                 Kobe International Conference Center, Kobe, Japan}

\begin{document}

\section{Introduction}

In a previous work\cite{Frison:2014esa} we have set a new strategy to improve the non-perturbative renormalisation (NPR) of the Kaon-Bag parameter $B_K$, a four-quark operator describing the neutral kaon oscillation. It included several new $\Nf=2+1+1$ and $\Nf=2+2$ ensembles with fine lattice spacing, in order to access high energy scale for the RI/SMOM schemes. Here we present a direct generalisation of this work to a basis of $5$ operators with mixing, those operators differing from the former one by their Dirac structure: 
\begin{eqnarray}
Q_1=&Q_{(27,1)}=&(\bar s\gamma_\mu d)(\bar s\gamma_\mu d)+(\bar s\gamma_\mu\gamma_5d)(\bar s\gamma_\mu\gamma_5d) \\
Q_2=&Q_{(8,8)_1}=&(\bar s\gamma_\mu d)(\bar s\gamma_\mu d)-(\bar s\gamma_\mu\gamma_5d)(\bar s\gamma_\mu\gamma_5d) \\
Q_3=&Q_{(8,8)_2}=&(\bar sd)(\bar sd)-(\bar s\gamma_5d)(\bar s\gamma_5d) \\
Q_4=&Q_{(6,6)_1}=&(\bar s\gamma_\mu d)(\bar s\gamma_\mu d)+(\bar s\gamma_\mu\gamma_5d)(\bar s\gamma_\mu\gamma_5d) \\
Q_5=&Q_{(6,6)_2}=&(\bar s\sigma_{\mu\nu} d)(\bar s\sigma_{\mu\nu} d)+(\bar s\sigma_{\mu\nu}\gamma_5d)(\bar s\sigma_{\mu\nu}\gamma_5d) .
\end{eqnarray}

Those operators allow for computations of the neutral kaon oscillation beyond the Standard Model (with anything other than a $V-A$ interaction), so that it is typically called ``SUSY basis'' \cite{Hudspith:2015,Boyle:2012qb}. But it also describes problems such as the $K\to\pi\pi$ $\Delta I=3/2$ decay. What it does not describe are things such as the $(8,1)$ sector of the $K\to\pi\pi$ $\Delta I=1/2$ decay, which is outside the scope of this work because of its very noisy disconnected diagrams. 

Compared to the $B_K$ study, this one will mainly consist in promoting most numbers to matrices. One major difference, though, is that we are only going to provide results from a single scheme ($\rismomgamma$), whereas the main result of the $B_K$ study was the better agreement of the $\rismomqslash$ scheme with perturbative running for that sector. Indeed, we are still assessing which definition of a SUSY $\rismomqslash$ scheme would be the best. 

In a first section I will first give some update about the generation of our ensembles and their scale setting, where some important efforts have been made. Then a second section will be dedicated to the fit model. And the last section will give both fit-independent and fit-dependent results.

\section{Ensembles and Scale-setting}

\begin{figure}[ht]
  \centering
  \includegraphics[width=0.35\linewidth]{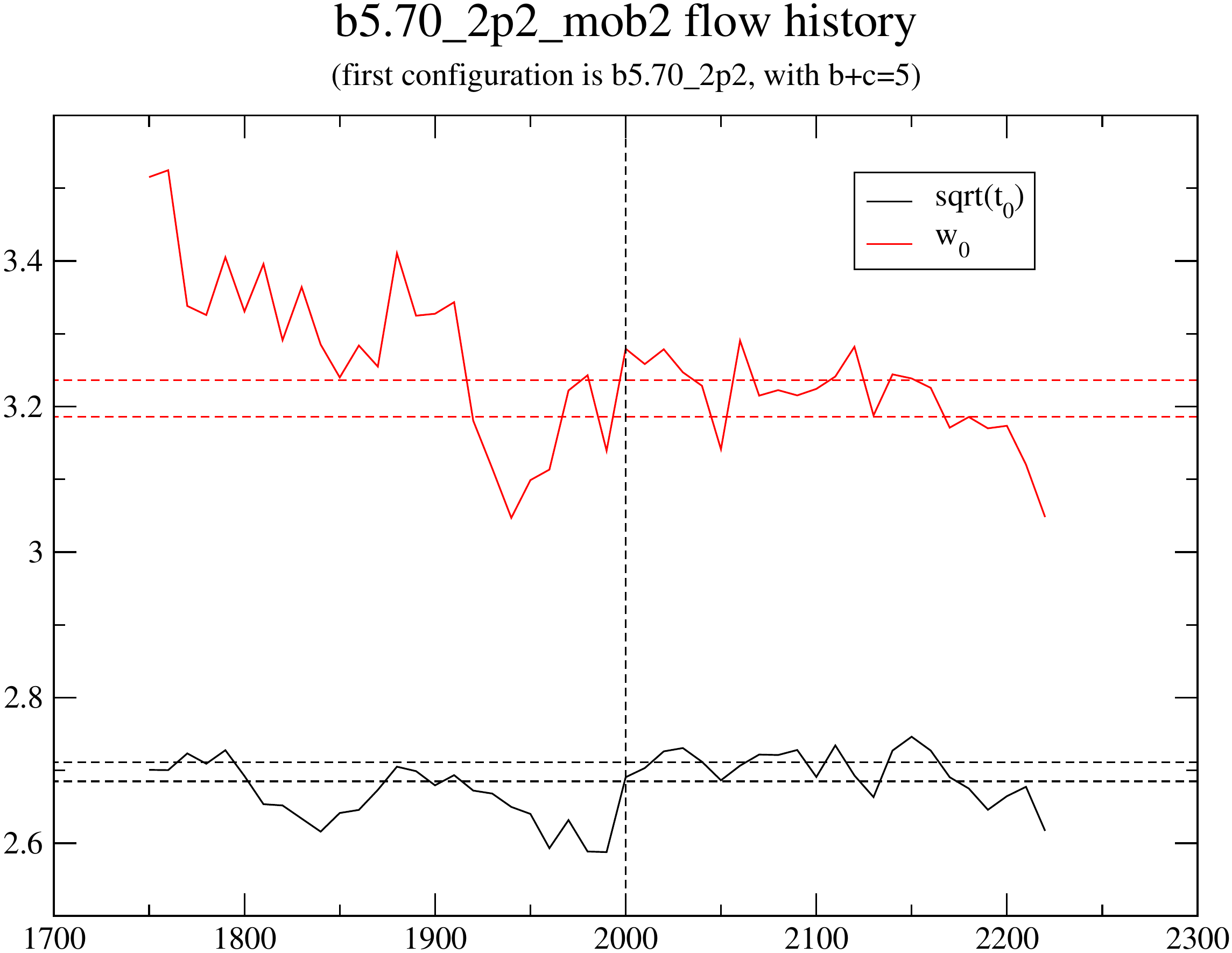}
  \hfill
  \includegraphics[width=0.35\linewidth]{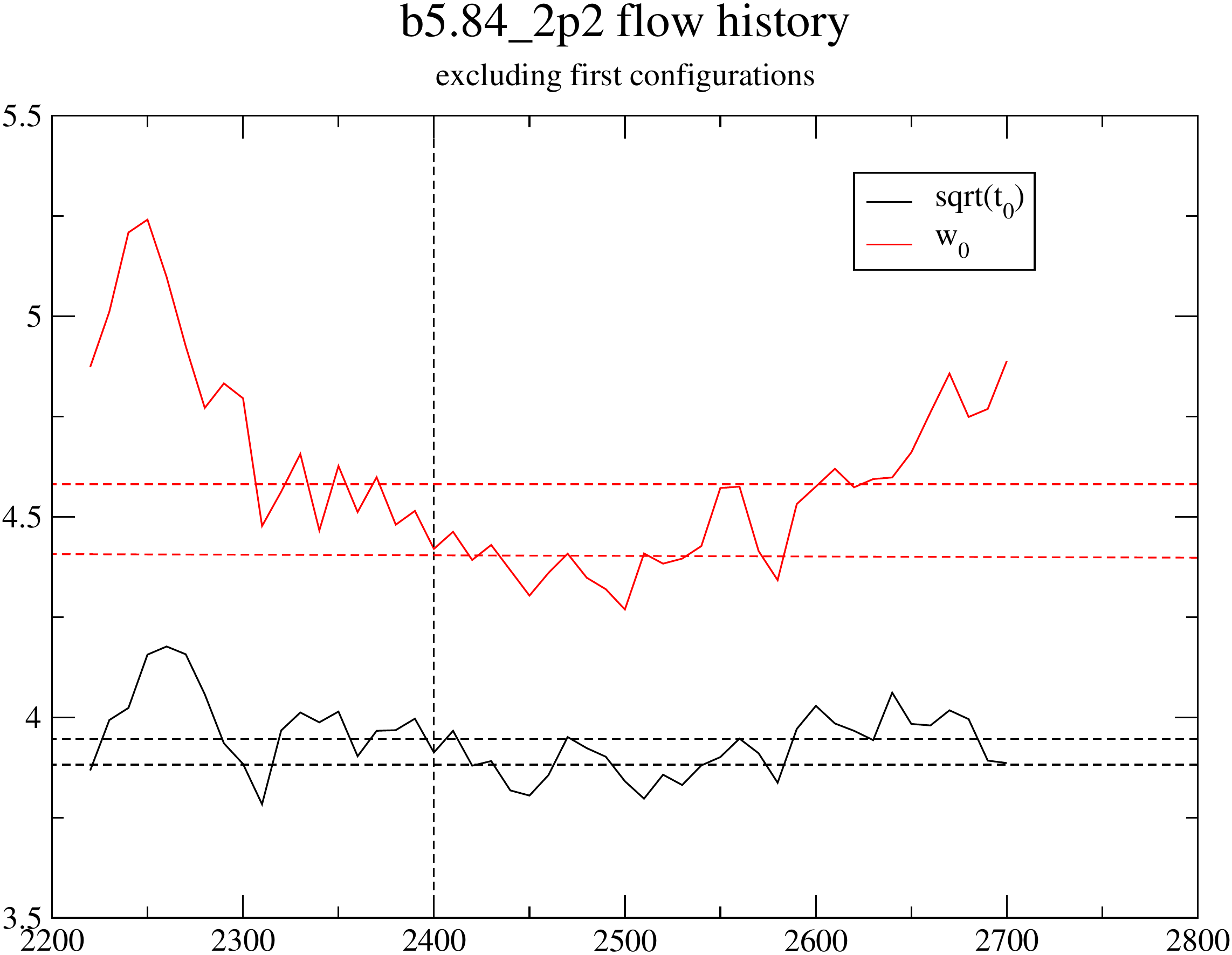}
  \caption{Wilson flow evolution for the two main ensembles. The vertical line shows where we start fitting, and the horizontal band shows the fit result. The x-axis shows the trajectory number.}
  \label{fig:flowhistory}
\end{figure}

\begin{figure}[ht]
  \centering
  \includegraphics[width=0.35\linewidth]{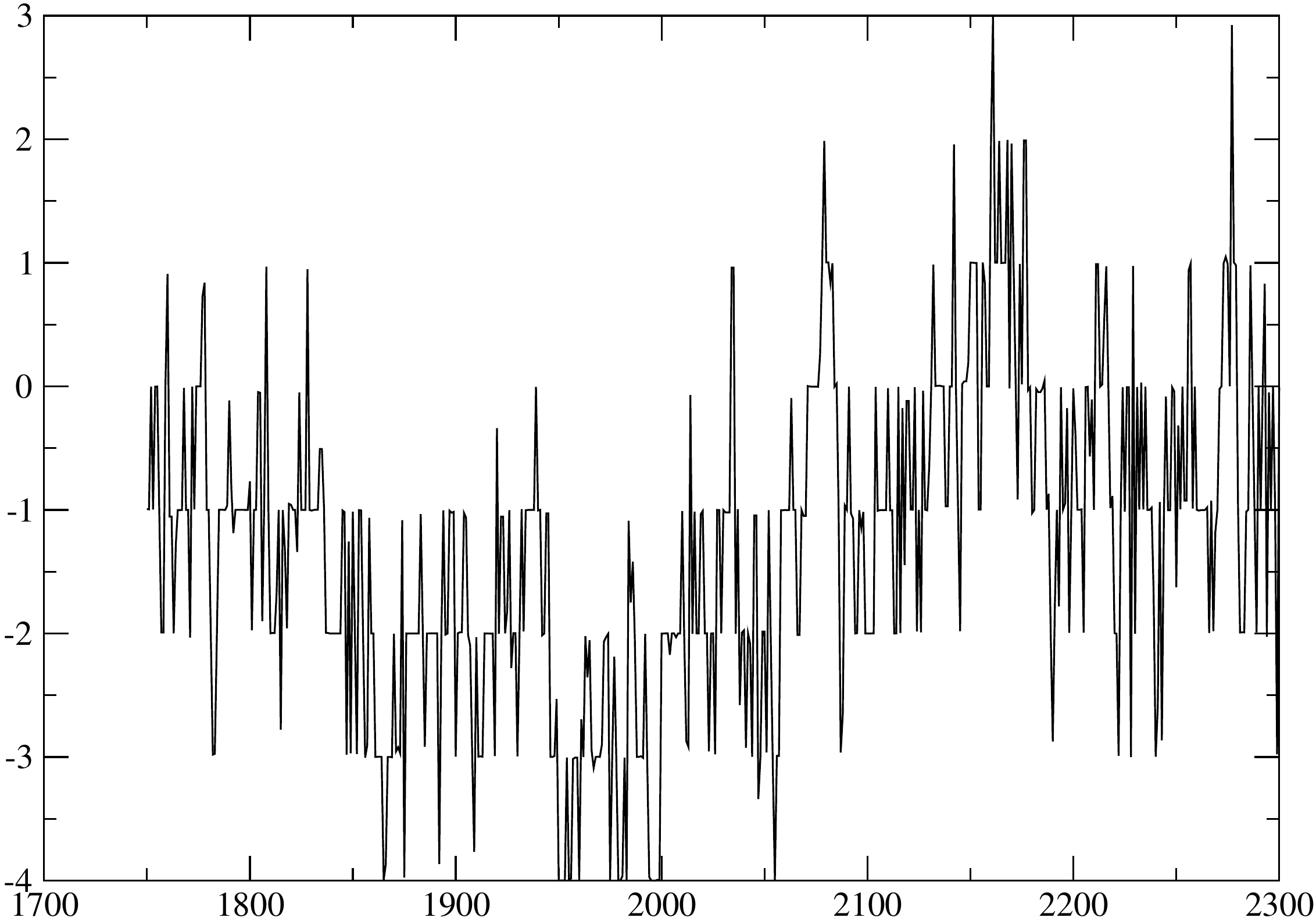}
  \hfill
  \includegraphics[width=0.35\linewidth]{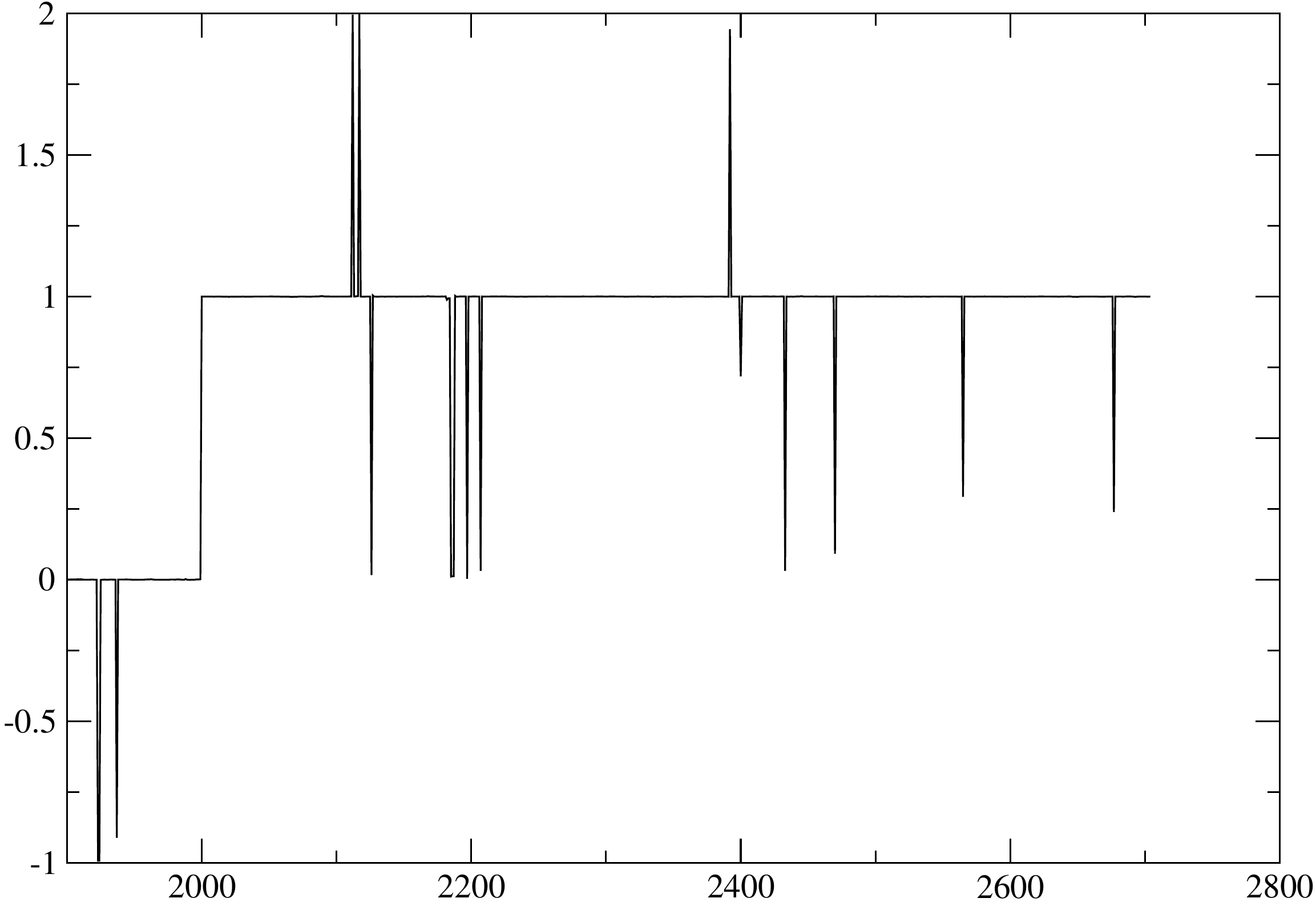}
  \caption{Topological charge for the two main ensembles, as an illustration of the progressive loss of tunnelling when we go beyond $a^{-1}=4\ \GeV$.}
  \label{fig:tcharge}
\end{figure}

The difficulty of generating fine ensembles has been widely experienced in the community. Although the NPR is a very robust quantity which is ideal for the first steps in the $\Nf=2+1+1$ world, we are hit by the problem of scale-setting. Indeed, a low-energy observable such as the Wilson flow is much more subject to long autocorrelations. Then, while a very limited number of configurations was enough to get small errors for the NPR Green functions, we felt necessary to generate many more configurations to make sure our Wilson flow was thermalised and ergodically sampled (see Figs.~\ref{fig:flowhistory}-\ref{fig:tcharge}).

Using the ensembles in Tab.~\ref{tab:ensembles} we concluded that the quark mass dependence (for both light and charm) is very small, and chose to neglect it in the present preliminary results, so that we only consider the $\Nf=2+2$ part of our step-scaling strategy\cite{Frison:2014esa}.

\begin{table}{ht}
\begin{center}
  \begin{tabular}{|c|c|c|c|c|c|}
    \hline
    $\beta$ & $L^3\times T\times L_5$ & $m_l$ & $m_s$ & $m_c$ & $a^{-1}$\\
    \hline
    $\mathbf{5.70}$ & $\mathbf{32^3\times 64\times 12}$ & $\mathbf{0.0047}$ & $\mathbf{0.0186}$ & $0.243$, $0.1$, $\mathbf{0.0186}$ & $\mathbf{3.0\ \textbf\GeV}$\\
    $5.70$ & $32^3\times 64\times 12$ & $0.002$ & $0.0186$ & $0.243$ & $3.0\ \GeV$\\
    $5.77$ & $32^3\times 64\times 12$ & $0.0044$ & $0.0166$ & $0.213$ & $3.6\ \GeV$\\
    $\mathbf{5.84}$ & $\mathbf{32^3\times 64\times 12}$ & $\mathbf{0.0041}$ & $\mathbf{0.0146}$ & $0.183$, $\mathbf{0.0146}$ & $\mathbf{4.3\ \textbf\GeV}$\\
    $5.84$ & $32^3\times 64\times 12$ & $0.002$ & $0.0146$ & $0.183$ & $4.3\ \GeV$\\
    \hline
  \end{tabular}
\end{center}
\caption{Fine $\Nf=2+1+1$ ensembles at our disposal. We highlight in boldface the ones actually used in this work. Those mass combinations have degenerate ``charm'' and ``strange'', so we call them $\Nf=2+2$ simulations.}
\label{tab:ensembles}
\end{table}


\section{Fit Models}

In a perturbative formulation, any lattice quantity naturally appears as 
\begin{equation}
F(p,a)=\sum_n g^n(a)F_n(ap).
\label{eq:latpert}
\end{equation}
As explained in \cite{Frison:2014esa}, one can 
therefore expect that at high-$p$ the most dramatic discretisation terms come as polynomials in $ap$, while the physical scale only appears in logarithms. When $p$ is large but $ap$ moderately small, one can design a double expansion and select a finite number of fit parameters. In our case, we found appropriate to describe discretisation effects by simply a term in $(ap)^2$ and a term in $(ap)^4$ (where the prefactors are matrices). In terms of the Green functions $\Lambda=Z^{-1}$ the fitting formula is
\begin{equation}
\Lambda(a_1,p_1)^{-1}\Lambda(a_2,p_2) = \left(1+A_1(a_1p_1)^2+B_1(a_1p_1)^4\right)^{-1}L_1^{-1} L_2\left(1+A_2(a_2p_2)^2+B_2(a_2p_2)^4\right), 
\end{equation}
for all momenta such that $p_1^2=p_2^2$. The $A_i$, $B_i$ and $L_i$ matrices are fit parameters. The indices run on any combination of ensemble and momentum orientation, so that for 2 ensembles and 2 orientations we have 4 ratios for each considered $p^2$, each ratio being a $5\times 5$ matrix. However, $A$ and $B$ only depend on the orientation while $L$ only depends on the ensemble. 

While each $\Lambda$ is $p$ dependent, those ratios are supposed to be $p$-independent (or even $1$ if the ensembles are the same), up to discretisation terms. All those ratios are therefore a rich signal of discretisation effects, signal which is already visible before any fit is performed. 

All-order perturbation theory gives a physical interpretation of why the fit works. However, an important property of our fitting formula is that it does not depend on the prior knowledge of perturbation theory at any given order. So the method is still purely non-perturbative. It is indeed much safer to study the convergence of perturbation theory independently from the continuum extrapolation.  

\section{Results}

On top of the various interesting ratios, we will present the results in terms of the previously studied step-scaling function\cite{Arthur:2010ht}. We also define RGI quantities: 
\begin{equation}
Z_{RGI} = C^T Z_{\rismom}
\end{equation}
and the (reduced) bilateral RGI step-scaling
\begin{equation}
\sigma_{RGI}(p,p_0) = C^T(p) \sigma_{\rismom}(p,p_0) C^{T-1}(p_0) \quad\text{and}\quad
\sigma'_s(p,p_0)=\sigma_s(p,p_0)-Id,
\end{equation}
where (using notations from \cite{Buras})
\begin{equation}
C = \left(1+\frac{\alpha}{4\pi}J\right) \alpha^{\gamma_0/2\beta_0} .
\end{equation}

\begin{figure}[ht]
  \centering
  \includegraphics[width=0.75\linewidth]{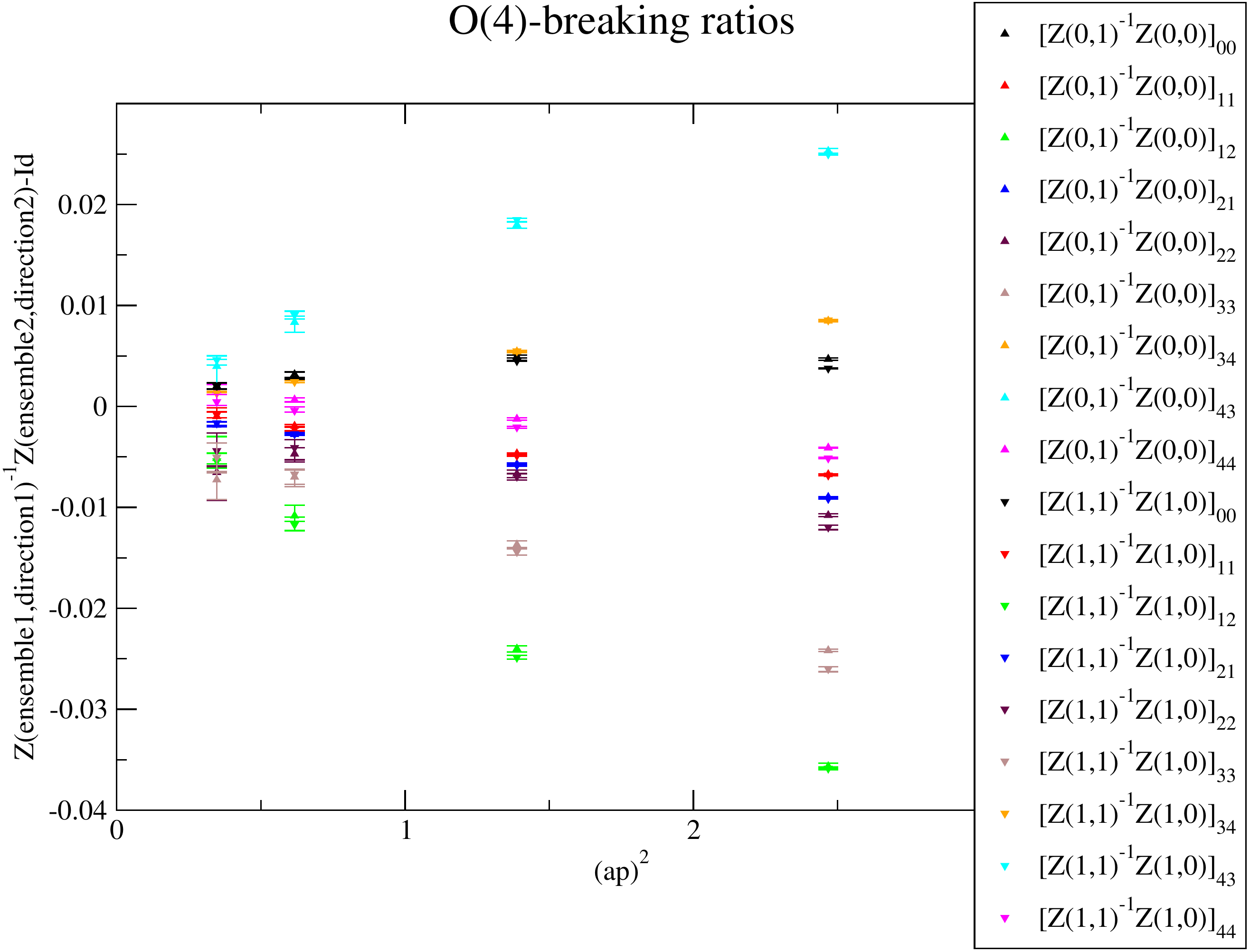}
  \caption{Subset of our ratios, where numerator and denominator use the same ensemble. We have only selected $p^2$ for which we have data for both directions at exactly the same $p^2$ without any interpolation, so that this signal is very reliable. We see that those quantities are surprisingly independent of $\beta$ (different curves of the same colour), even though we do see some discrepancy at our tiny statistical level for 4 of the 5 diagonal elements.}
  \label{fig:ratios}
\end{figure}

\begin{figure}[ht]
  \centering
  \includegraphics[width=0.45\linewidth]{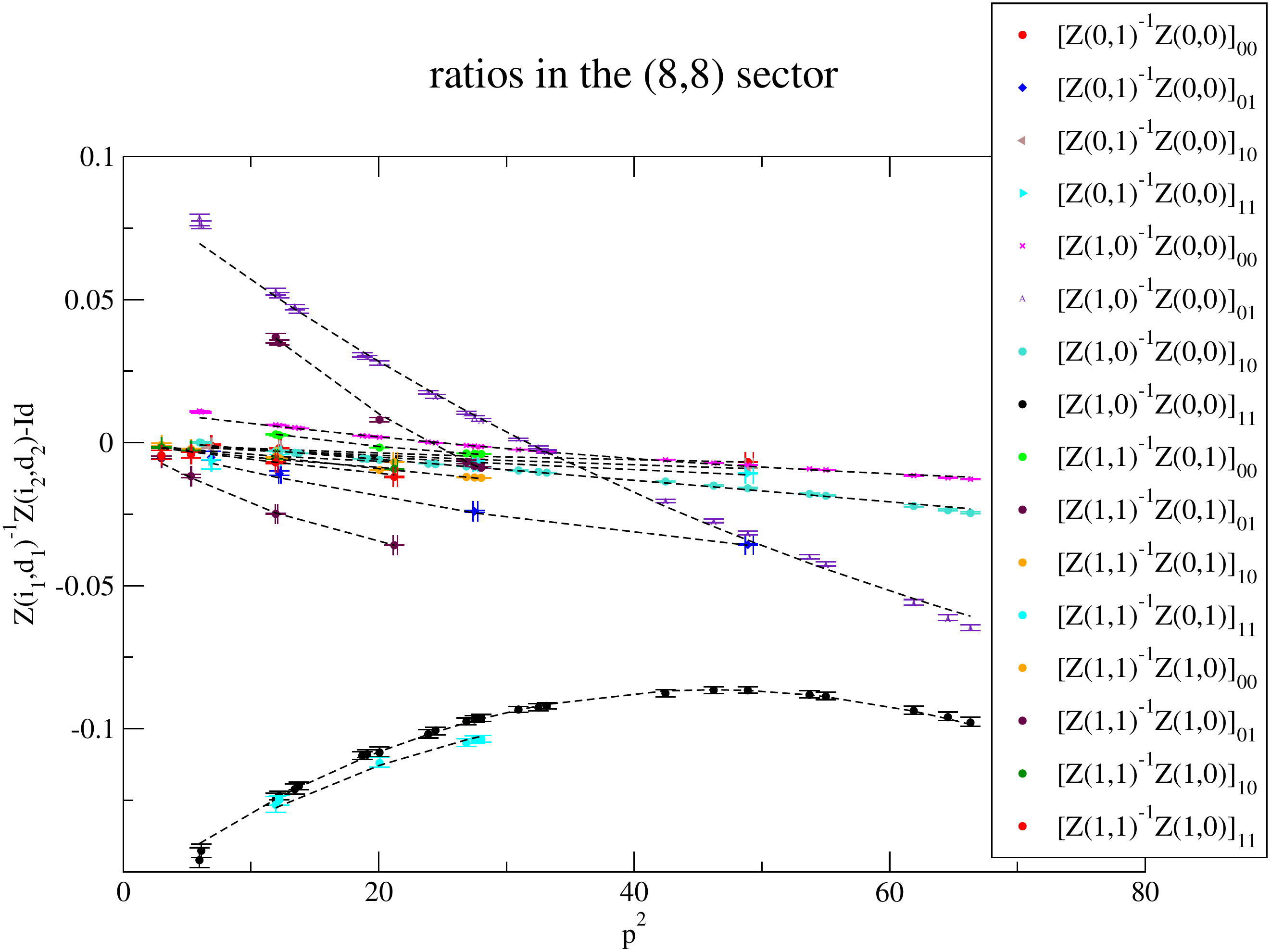}
  \hfill
  \includegraphics[width=0.45\linewidth]{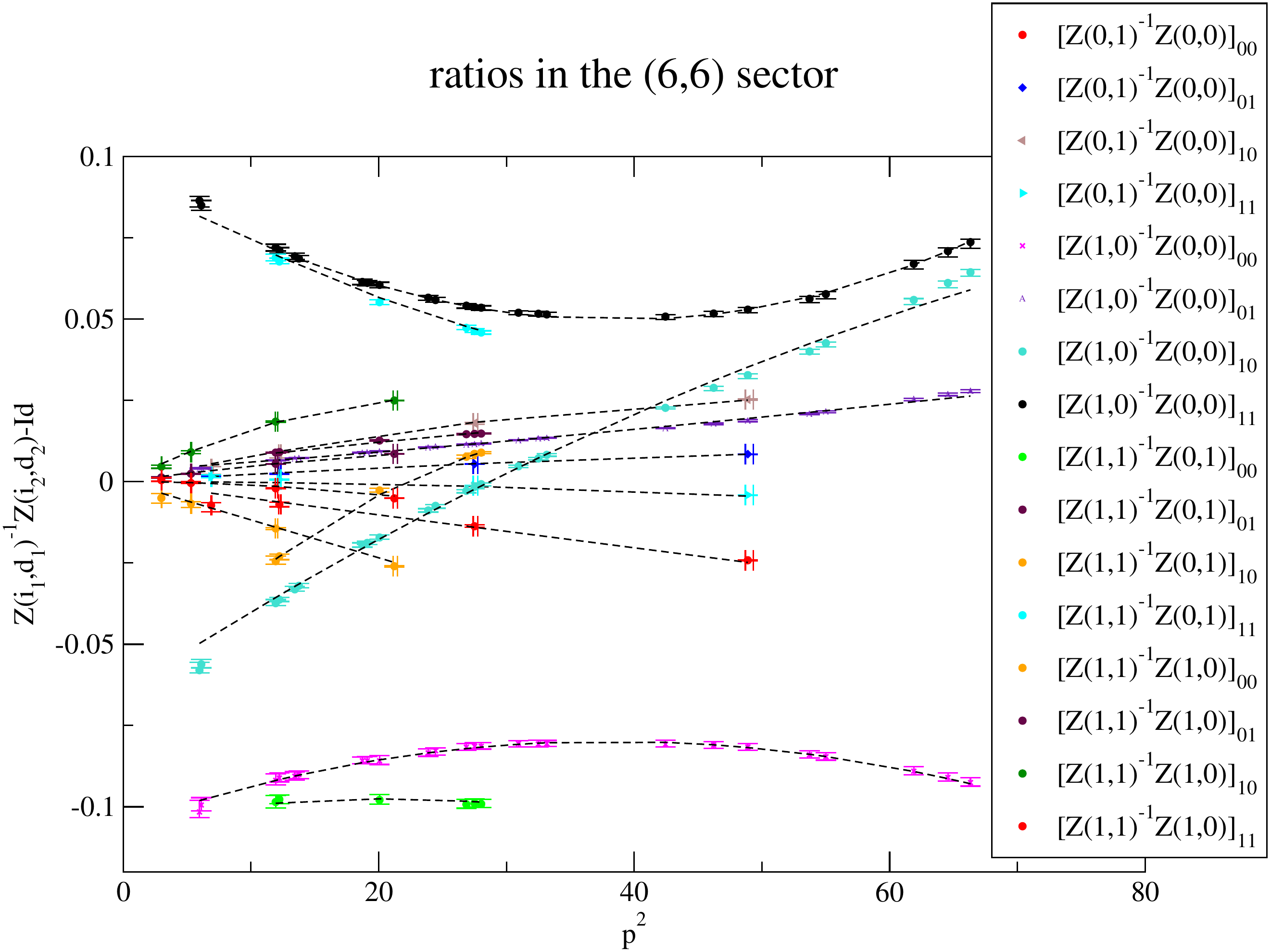}
  \caption{Pushing to a larger $(ap)^2$ cut-off than the one chosen in the final analysis, one can clearly see the curvature from $(ap)^4$ terms. Moreover, most of the data is still in striking agreement with the fit.}
  \label{fig:66ratios}
\end{figure}

\begin{figure}[ht]
  \centering
  \includegraphics[width=0.75\linewidth]{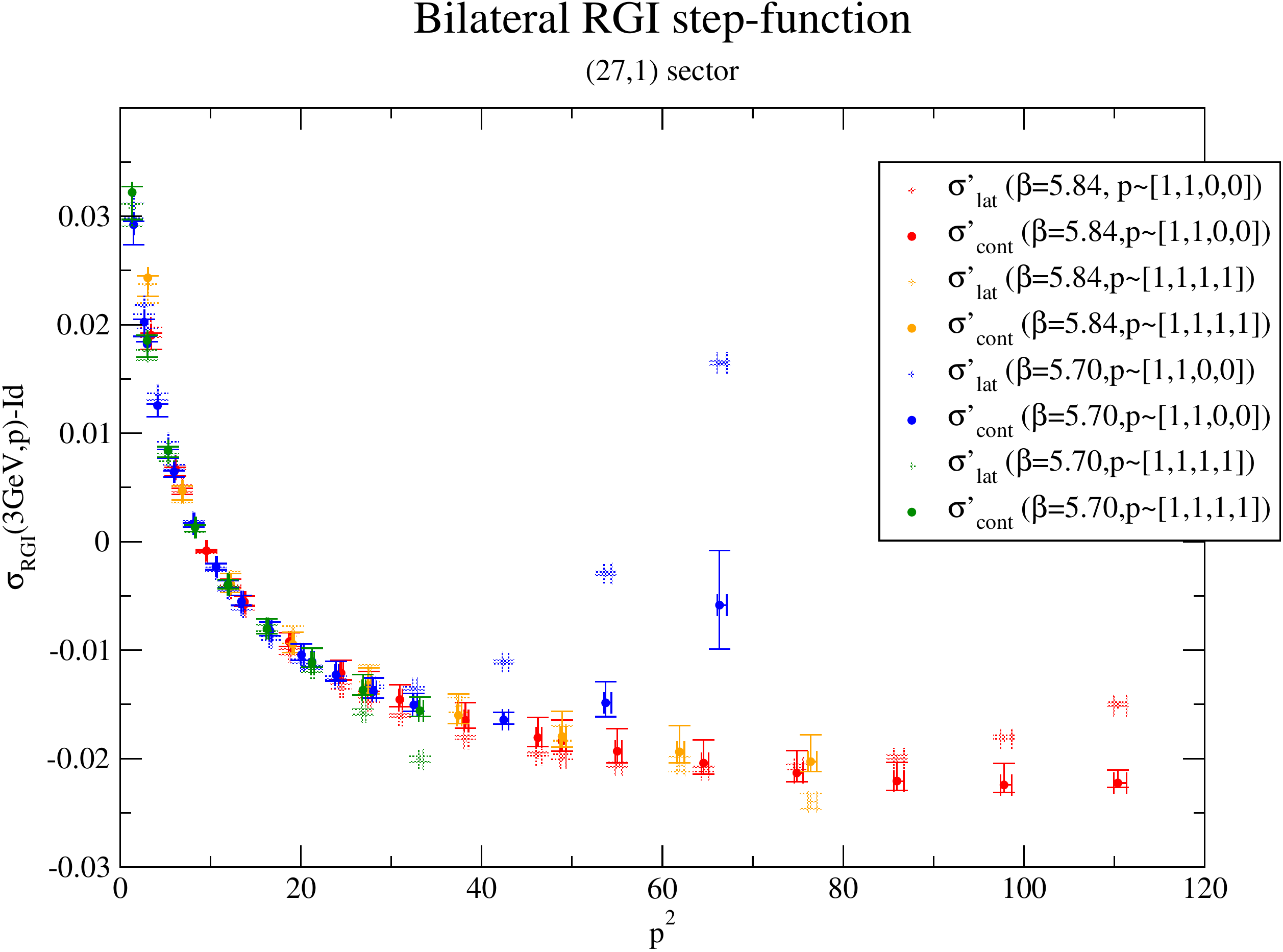}
  \caption{The reduced bilateral RGI step-function would have all its elements to zero if perturbation theory were perfect. We plot for each ensemble and orientation both the raw data $\sigma'_{lat}$ and the corrected data $\sigma'_{cont}$ whose main part of the discretisation effects is removed thanks to the results of the global fit. For $(27,1)$ we seem to be approaching a plateau, and could gain up to $2\%$ in truncation error. But having $\rismomqslash$ was decisive to bring a solid conclusion in \cite{Frison:2014esa}, because they both agreed well at large $p^2$. }
  \label{fig:sRGI271}
\end{figure}

\begin{figure}[ht]
  \centering
  \includegraphics[width=0.45\linewidth]{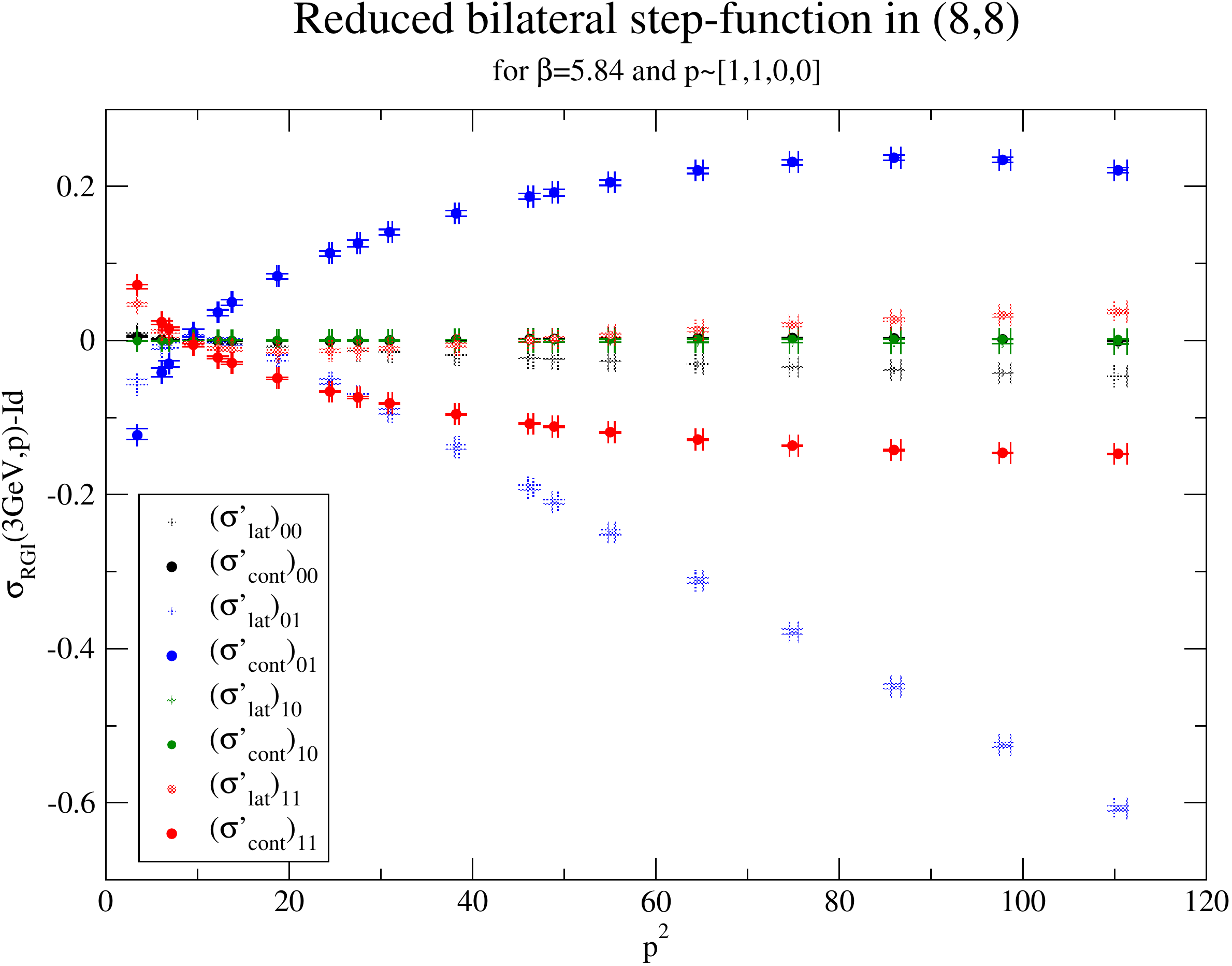}
  \hfill
  \includegraphics[width=0.45\linewidth]{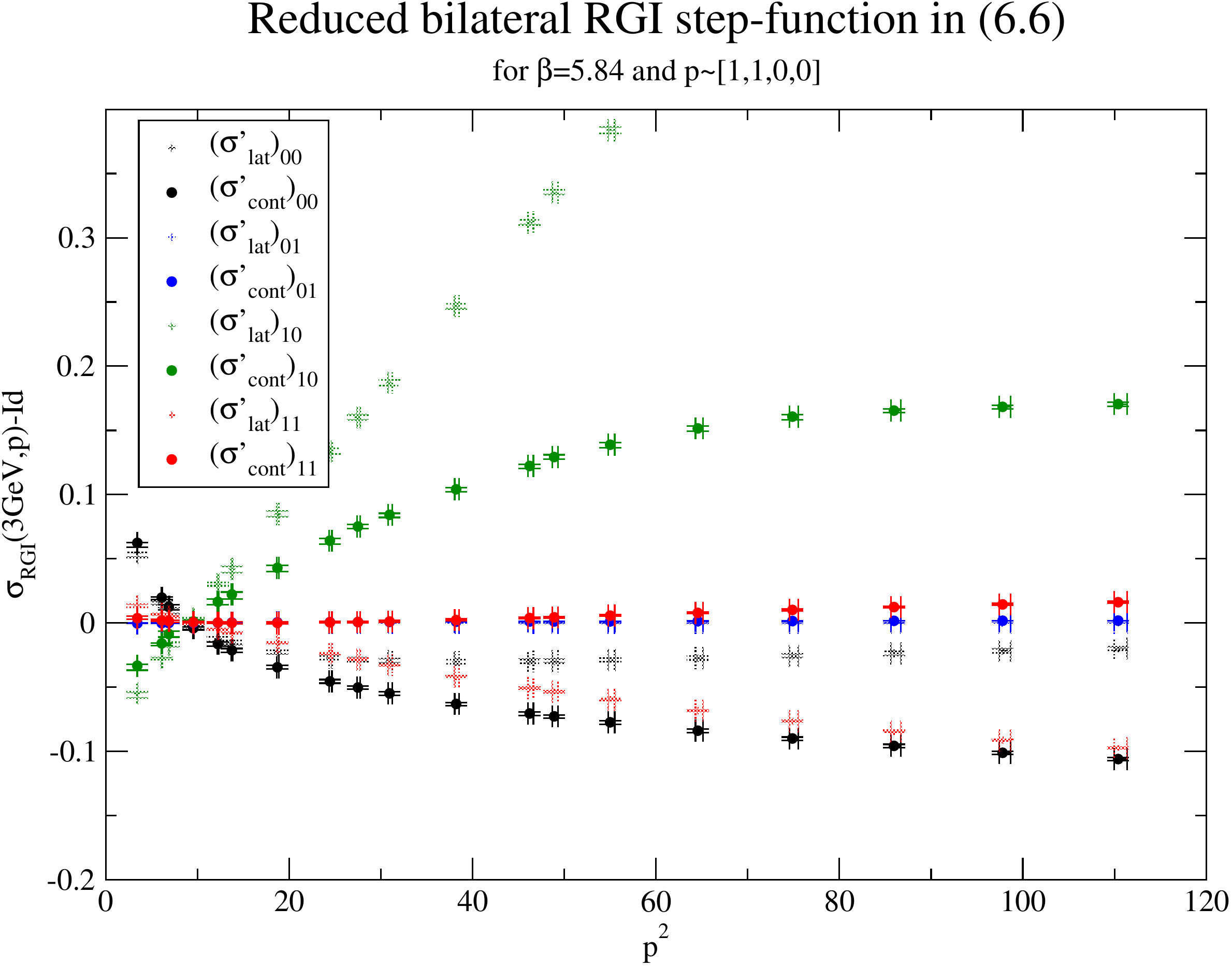}
  \caption{Reduced bilateral RGI step function in the BSM sectors. Thanks to the subtraction of discretisation effects, the main contributions come closer to a plateau. This scale dependence is certainly indicative of large perturbative truncation error in SUSY $B_K$ operators renormalised at $3\ \GeV$ in the $\rismomgamma$ scheme. Here increasing the scale has a much more dramatic effect than in the $B_K$ sector, up to $\sim 20\%$.}
  \label{fig:DRGI44}
\end{figure}

Because we do not want to depend on the fit as a black box, we start by looking at the $O(4)$-breaking part of the discretisation errors (Fig.~\ref{fig:ratios}), which can be studied separately for every ensemble. The fact the $\beta$ dependence is hardly visible is already a strong argument in favour of our fit model, showing the weak influence of logarithic terms in Eq.~(\ref{eq:latpert}). Then we can look at the full set of ratios in the new $(8,8)$ and $(6,6)$ sectors, compared with the result of the fit (Fig.~\ref{fig:66ratios}). Finally in Figs.~\ref{fig:sRGI271}-\ref{fig:DRGI44} we use our model to correct the data.

Without the comparison with a second scheme, we do not really know which $p^2$ is a good compromise between small discretisation errors and small truncation errors. While we might want to push harder in the future, we choose to stay on the safe side and give a preliminary result at $5\ \GeV$ (central value is taken from the corrected data of the finest ensemble, while the second number in parenthesis is an estimate of the systematic error given by the full difference between the fine ensemble and the coarse ensemble): 
\begin{equation}
\sigma_{\text{\tiny{\rismom}}}(3\ \GeV,5\ \GeV) = 
\left(
\begin{tabular}{ccccc}
1.0248(5)(7) & 0 & 0 & 0 & 0 \\
0 & 1.0239(5)(5) & 0.168(3)(2) & 0 & 0\\
0 & 0.0064(4)(2) & 0.764(4)(5) & 0 & 0\\
0 & 0 & 0 & 0.838(3)(4) & 0.0002(4)(6)\\
0 & 0 & 0 & -0.138(2)(5) & 1.142(1)(1)\\
\end{tabular}
\right)
.\nonumber
\end{equation}
Please do not forget that some systematics have been neglected, mainly the mass dependences, and they are expected to lie at the per mile level, based on past evidence with non-exceptional schemes. 


\section{Conclusion}

We have shown that there is a strong relation between the $p$-dependence and the $a$-dependence of $\rismom$ renormalisation factors, and that we are able to exploit it, even in cases with important operator-mixing. While the discretisation effects can be large, this technique allows to keep them under control up to very high scales, in that a parametric form describes the data well and can be subtracted, without compromising the renormalisation's non-perturbative character. 

However, despite very encouraging results, we still have not reached a scale at which perturbation theory fully agrees with the non-perturbative results. For several matrix elements it is even hard to tell whether we are really getting closer to a plateau or not. Comparing these results with a $\rismomqslash$ scheme, as we did for $B_K$, would certainly illuminate our interpretation, since the RGI result should be independent of intermediate scales and schemes if perturbation theory is accurate. It is an objective for the near future. 

On a longer term we would certainly like to generate a new $\Nf=2+2$ ensemble at an intermediate lattice spacing, and more configurations for some of the ensembles we have not used here. But a reliable tunnelling is expensive, so a full treatment of every systematic is unlikely to enter the scope of this exploratory project before a massive algorithmic advance appears.


\begin{thebibliography}{99}

\bibitem{Frison:2014esa} 
  J.~Frison, P.~Boyle and N.~Garron,
  PoS LATTICE {\bf 2014}, 285 (2015)
  [arXiv:1412.0834 [hep-lat]].

\bibitem{Hudspith:2015} 
  R.~Hudspith et al,
  PoS LATTICE {\bf 2015}, 327 (2015)

\bibitem{Boyle:2012qb} 
  P.~A.~Boyle {\it et al.} [RBC and UKQCD Collaborations],
  Phys.\ Rev.\ D {\bf 86}, 054028 (2012)
  [arXiv:1206.5737 [hep-lat]].


\bibitem{Buras}
  A.~J.~Buras, S.~Jager and J.~Urban,
  Nucl.\ Phys.\ B {\bf 605}, 600 (2001)
  [hep-ph/0102316].


\bibitem{Arthur:2010ht} 
  R.~Arthur {\it et al.}  
  Phys.\ Rev.\ D {\bf 83}, 114511 (2011)
  [arXiv:1006.0422 [hep-lat]].

\end{thebibliography}
\end{document}